\documentclass[prl,aps,twocolumn,superscriptaddress]{revtex4-1}
\usepackage{amssymb,amsfonts,amsmath,graphicx,color,mathptm,times}

\def\bra#1{\langle{#1}\vert}
\def\ket#1{\vert{#1}\rangle}
\def\braket#1{\langle{#1}\rangle}

{\catcode`\|=\active 
 \gdef\Braket#1{\begingroup
\mathcode`\|32768\let\vert\BraVert\left<{#1}\right>\endgroup}}
\def\BraVert{\egroup\,\mid\,\bgroup}

\def\tr{\mbox{tr}}
\newcommand{\ent}{S}
\newcommand{\q}{{Q}}
\newcommand{\s}{\mathcal{S}}
\newcommand{\e}{\mathcal{E}}
\newcommand{\se}{\mathcal{SE}}

\newcommand{\A}{\hat{\bf A}}
\newcommand{\M}{\hat{\bf M}}

\definecolor{Blue}{rgb}{0,0,1}
\definecolor{Red}{rgb}{0,0,0}
\definecolor{Green}{rgb}{0,1,0}
\definecolor{Purp}{rgb}{.2,0,.2}
\definecolor{white}{rgb}{1,1,1}

\begin{document}
\title{A non-equilibrium quantum Landauer principle}

\author{John Goold}
\email{jgoold@ictp.it}
\affiliation{The Abdus Salam International Centre for Theoretical Physics (ICTP), Trieste, Italy}

\author{Mauro Paternostro}
\email{m.paternostro@qub.ac.uk}
\affiliation{Centre for Theoretical Atomic, Molecular and Optical Physics, School of Mathematics and Physics, Queen's University, Belfast BT7 1NN, United Kingdom}

\author{Kavan Modi}
\email{kavan.modi@monash.edu}
\affiliation{School of Physics, Monash University, Victoria 3800, Australia}

\begin{abstract}
Using the operational framework of completely positive, trace preserving operations and thermodynamic fluctuation relations, we derive a lower bound for the heat exchange in a Landauer erasure process on a quantum system. Our bound comes from a non-phenomenological derivation of the Landauer principle which holds for generic non-equilibrium dynamics. Furthermore the bound depends on the non-unitality of dynamics, giving it a physical significance that differs from other derivations. We apply our framework to the model of a spin-1/2 system coupled to an interacting spin chain at finite temperature.
\end{abstract}

\date{\today}

\maketitle

{\it Introduction.---}
The most convincing evidence that information is indeed physical is provided by Landauer's principle~\cite{landauer}. It states that the logically-irreversible erasure of information carried by a physical system comes at the expense of heat dissipation to the environment. Stated equivalently, the principle provides a direct link between the domains of information theory and thermodynamics. The deep consequences of Landauer's principle were instrumental for Bennett to attach a minimum entropy production to the logically irreversible procedure of erasure~\cite{bennett, plenio}, thus operating an information-theoretical exorcism of Maxwell demon and recognising that computation can be done reversibly, in principle, requiring no heat production.

Turning to quantum systems, it is surprising that very few papers have a clear operational framework which is suitable to understand the emergence of Landauer's principle from the underlying microscopic equations. In a recent work~\cite{reeb}, Reeb and Wolf use techniques from quantum statistical mechanics to prove that a finite-size environment can provide a tighter bound to the quantity of heat generated in an erasure process. The usual Landauer bound
\begin{gather}\label{landauer}
\beta\braket{\q}\ge \Delta \ent
\end{gather}
is retrieved when the thermodynamic limit is taken in the environment. In Eq.~\eqref{landauer}, $\braket{\q}$ is the average heat exchange with the bath and $\Delta \ent$ is the information theoretic entropy change. The finite size corrections to Eq.~\eqref{landauer} proposed in Ref.~\cite{reeb} are, in some sense, suggestive of intrinsic non-equilibrium dynamics that we would expect away from the thermodynamic limit and have been explored previously in the context of irreversible entropy production \cite{esposito1}.

One way to describe the thermodynamics of systems where thermal and quantum fluctuations cannot be ignored is to consider the so called quantum fluctuation relations~\cite{jrev, Jarzynski:97, jarzynski:04, Crooks}. The fluctuation relations, extended to the quantum mechanical domain~\cite{Tasaki, mrev} are a promising route to understand the thermodynamics of small quantum systems that are operating under non-equilibrium conditions. Crucially, recent work has demonstrated that this formalism is a tangible route for the experimental exploration of quantum thermodynamics~\cite{dorner2, mauro, nmr}.

In this Letter we bring together the tools of open quantum systems, non-equilibrium statistical mechanics and quantum information theory to derive a non-phenomenological lower bound for heat generated in an erasure process. We begin by recasting the erasure protocol given in Ref.~\cite{reeb} from the point of view of fluctuation relations. Extension of the fluctuation relation formalism to the open quantum-system framework leads to difficulties unless fairly restrictive assumptions are made~\cite{esposito, lindblad}. A series of papers have attempted to derive fluctuation relations from the operational viewpoint, employing the full machinery of completely positive, trace preserving operations~\cite{vedral, kafri, rastegin, albash, rastegin2, johnkavan2}, which are ubiquitous in quantum information and it was found that fluctuation-like relations can hold if the open system evolution is unital~\cite{petruccione}.

We construct a distribution of heat dissipated by an erasure protocol involving a finite size environment interacting with a quantum system. We show that the non-unitality of a given open-system dynamics can lead to a tighter bound on the average heat exchanged during the process than previously known bounds. In addition, and more importantly, the methodology developed here paves the way to the tantalising possibility of analysing the thermodynamics of computation under non-equilibrium conditions. Our work brings together concepts from several disciplines of physics and constitutes a promising route for the construction of a formal platform for exploring the efficiency and limitations of small-scale thermo-devices operating at and well within the quantum mechanical domain~\cite{kosloff}.

{\it Erasure protocol.---}
Starting from the analysis put forward in Ref.~\cite{reeb}, the starting point of our investigation is embodied by a general erasure protocol that satisfies the following set of pre-requisites:
\begin{itemize}
\item[1.] A system $\s$, whose information content we want to erase, is subjected to an environment-aided erasure protocol. We call $\hat H_\s$ the free Hamiltonian of the system.
\item[2.] We label such environment as $\e$ and assume that the initial total $\se$ state is fully uncorrelated, i.e., $\hat\rho_{\se}=\hat\rho_\s \otimes \hat\rho_\e$.
\item[3.] The environment is initially in the Gibbs state $\hat\rho_\e = e^{-\beta \hat H_\e}/{\cal Z}_\e$, with Hamiltonian $\hat H_\e = \sum_m E_m \ket{r_m}\bra{r_m}$, the inverse temperature $\beta$, and the partition function ${\cal Z}_\e=\tr[e^{-\beta\hat H_\e}]$.
\item[4.] The system and environment interact via a perfectly unitary mechanism described by the time propagator $\hat U$ generated by the total Hamiltonian $\hat H=\hat H_\s+\hat H_\e+\hat H_{\se}$.
\end{itemize}
These points constitute a non-restrictive set of {\it rules} for the erasure protocol to be performed. 


{\it Thermodynamics of the environment.---}
We describe the erasure process as the protocol that, starting from the joint $\se$ initial state, generates the following reduced state of the environment only
\begin{gather}
\hat\rho'_\e = \tr_\s [\hat U(\hat\rho_\s \otimes \hat\rho_\e)\hat U^\dag]
= \sum_{l} \hat A_{l} \, \hat\rho_\e \, \hat A_{l}^\dag.
\label{operation}
\end{gather}
Here, $\hat A_{l = jk} = \sqrt{\lambda_j} \braket{s_k \vert  \hat U \vert  s_j}$ where $\{\lambda_j\}$ and $\{\ket{s_j}\}$ are the eigenvalues and eigenstates of $\hat\rho_\s$. It can be rigorously shown that the operators $\hat A_l$ satisfy the trace-preserving condition $\sum_l \hat A^\dag_l \hat A_l = \openone_\e$. It is worth stressing that Eq.~\eqref{operation} does not embody a map, but an operation. In fact, we can vary $\hat\rho_\s$, thus changing the form of the $\hat A_l$'s while keeping $\hat\rho_\e$ fixed. 

In analogy to the work distribution, we now define the heat distribution for the environment~\cite{johnkavan} as
\begin{gather}\label{Qdist}
P(\q) = \sum_{l,m,n} \braket{r_n \vert \hat A_l \vert r_m} (\hat\rho_\e)_{mm}\braket{r_m \vert \hat A^\dag_l \vert r_n} \delta(\q-E_{nm})
\end{gather}
with $(\hat\rho_\e)_{nm}=\bra{r_n}\hat\rho_\e\ket{r_m}$ the matrix element of the environmental initial state in the basis of its eigenstates and $E_{nm}=E_n-E_m$. The first moment of the heat distribution is the average heat, which can be written as 
\begin{align}
\braket{\q} &= \int \q \, P(\q) \, d\q = \sum_{n} E_n (\hat\rho'_\e)_{nn} - \sum_{m} E_m\sum_{l} \tr[\hat A_l \hat\rho_\e \hat A^\dag_l] \notag\\
&= \tr[\hat H_\e \hat\rho'_\e]- \tr[\hat H_\e \hat\rho_\e].
\end{align}
The distribution of heat contains much more information than just the first moment. For instance, Jarzynski has found an important equality using the work distribution \cite{Jarzynski:97}, which relates the average exponentiated work to the equilibrium free energy. In the same spirit, we now evaluate the average exponentiated heat to derive a heat fluctuation relation. We have
\begin{align}
\braket{e^{-\beta \q}} =& \int e^{-\beta \q} \, d\q \, P(\q) \notag\\
=&\sum_{l,m,n} \braket{r_n \vert \hat A_l \vert r_m} \braket{r_m \vert \hat A_l^\dag \vert r_n}(\hat\rho_\e)_{mm} e^{-\beta E_{nm}}.
\end{align}

Plugging in $e^{-\beta E_m}/\cal{Z}_\e$ for $(\hat\rho_\e)_{mm}$ and summing over $n$ we have
\begin{gather}
\braket{e^{-\beta \q}} = \sum_{l} \tr[\hat  A_l^\dag \hat\rho_\e \hat A_l]
=\tr[\A \, \hat\rho_\e ]\label{expoA}
\end{gather}
with $\A=\sum_{l} \hat A_l \hat A_l^\dag$. On the other hand if we expand the operator $\A$ in terms of the initial states of both system and environment under the action of the unitary process $\hat U$ and use the cyclicity of trace we get
\begin{gather}
\braket{e^{-\beta \q}} =\tr[\hat\rho_\s \otimes \openone_\e \, \hat U^\dag \, \openone_\s \otimes \hat\rho_\e \, \hat U]
=\tr[\M \, \hat\rho_\s],\label{expoM}
\end{gather}
where $\M = \tr_\e[\hat U^\dag \, \openone_\s \otimes \hat\rho_\e \, \hat U]$. 

Eqs.~\eqref{expoA} and \eqref{expoM} are the central results of this Letter. We will discuss their significance below. However, let us first obtain a lower bound on the average heat exchange. In order to formulate such a bound from the above equality, we use the Jensen inequality $\braket{f(x)} \ge f(\braket{x})$, which holds for any convex function $f(x)$. That is, using $\braket{e^{-\beta \q}}\ge e^{-\beta \braket{\q}}$ we have the desired bound
\begin{gather}\label{boundtotal}
\beta\braket{\q}\ge{\cal B}_\q,
\end{gather}
where ${\cal B}_\q =-\ln(\tr[\A \, \hat\rho_\e]) = -\ln(\tr[\M \, \hat\rho_\s ])$. 

\begin{figure*}[t]
\center{{\bf (a)}\hskip7cm{\bf (b)}\hskip4cm{\bf (c)}}
\includegraphics[width=0.85\columnwidth]{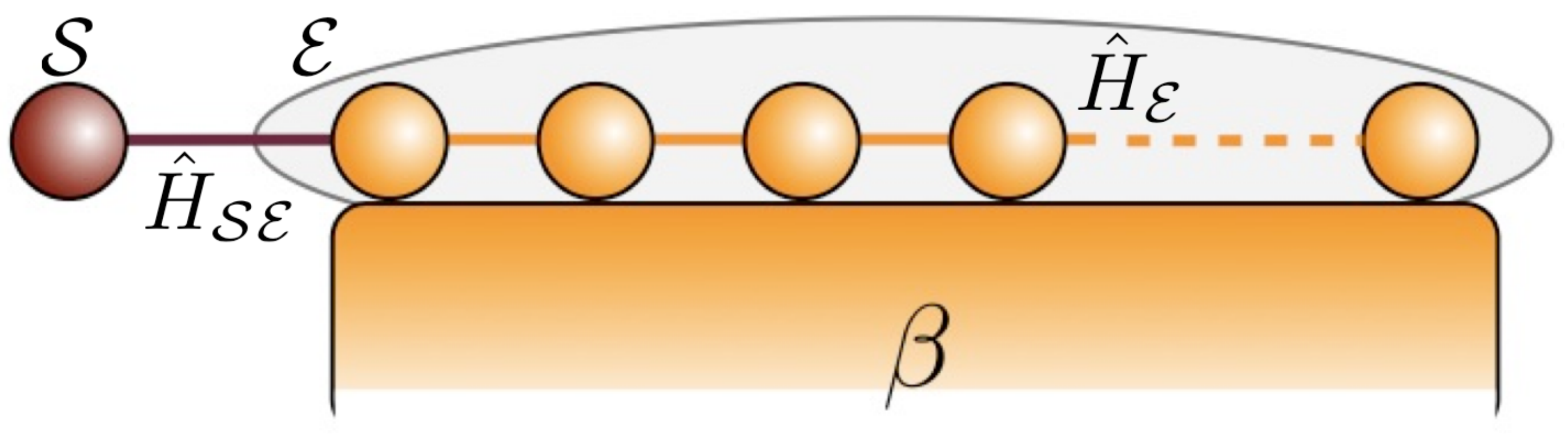}~~\includegraphics[width=0.6\columnwidth]{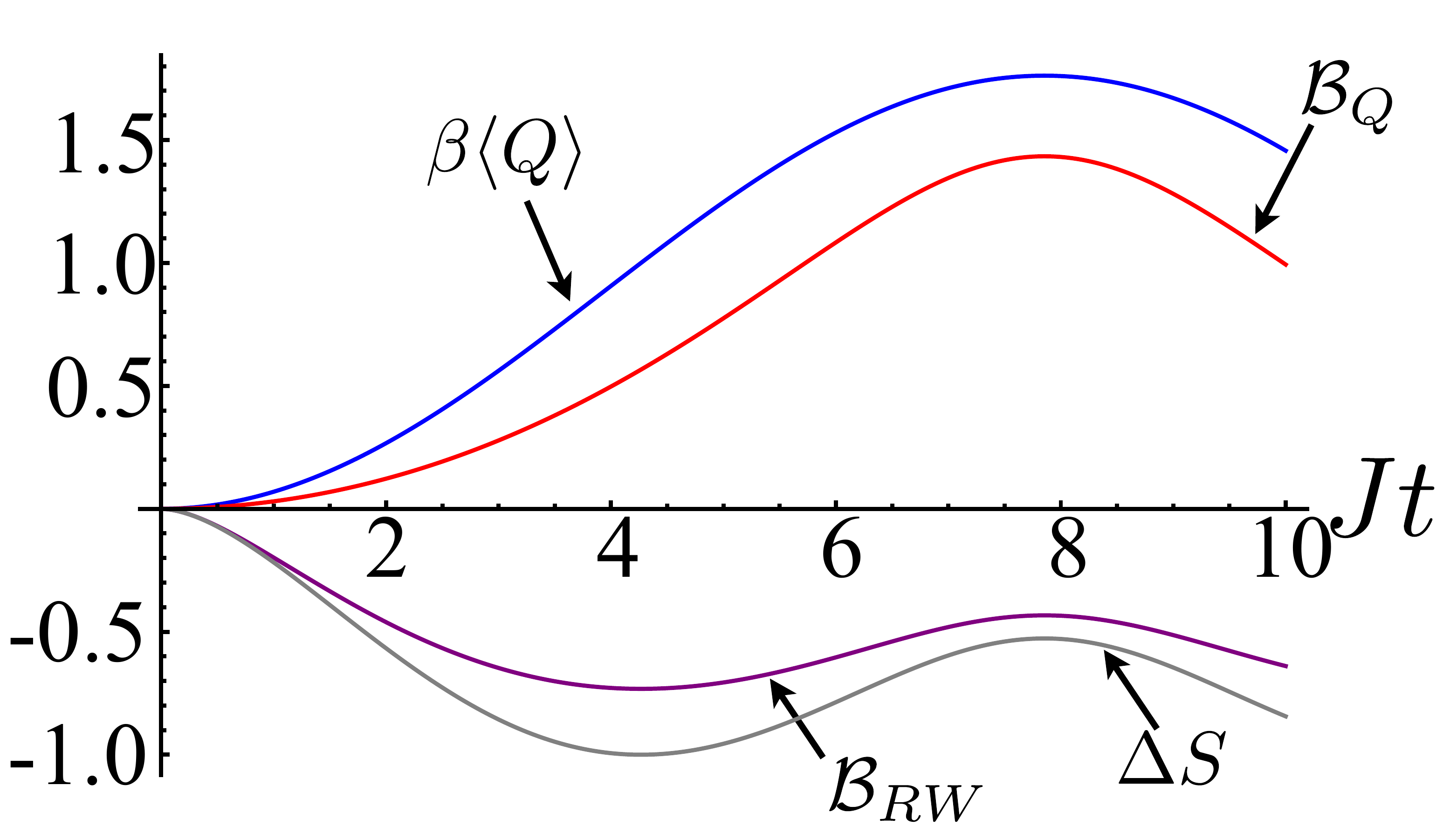}~~\includegraphics[width=0.55\columnwidth]{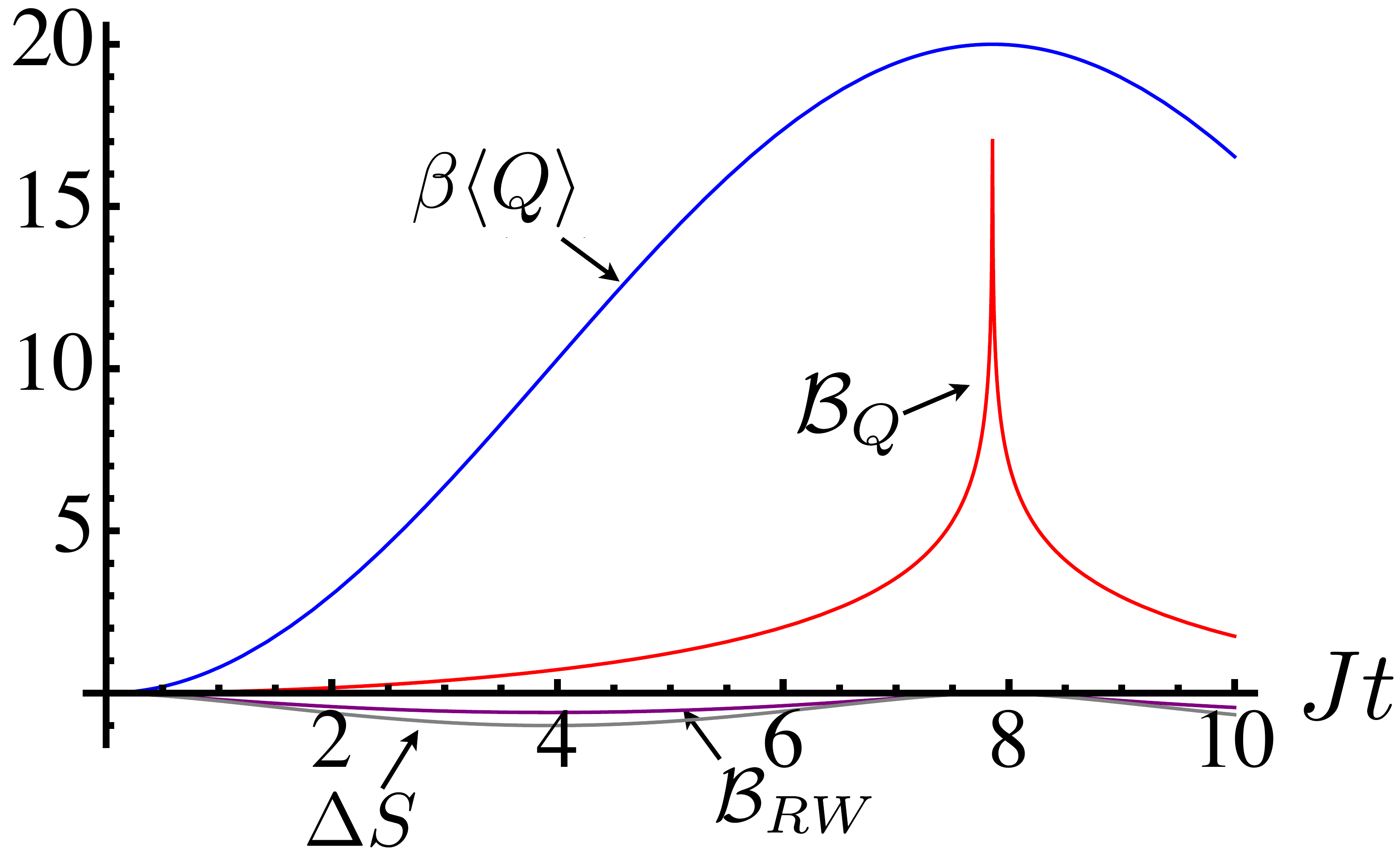}
\caption{{\bf (a)} Schematic representing the system under consideration. {\bf (b)} Comparison between $\beta\braket{\q}$, the bound ${\cal B}_\q$ derived in Eq.~\eqref{boundtotal}, and the one found in Ref.~\cite{reeb} for a spin-1/2 particle interacting for a dimensionless time $Jt$ with a single-spin environment at inverse temperature $\beta=1$. We also plot the change in entropy $\Delta \ent$. All the quantities are studied against the initial preparation $\alpha\ket{1}_\s+\sqrt{1-\alpha^2}\ket{0}_\s~(\alpha\in{\mathbb R})$ of the system state. {\bf (c)} Analogous comparison as in panel {\bf (b)}, but performed for $\beta=10$. In both panels we have used $B_0/J=B/J=1$ and $J_0/J=0.1$.}\label{spinenvironment}
\end{figure*}

Care should be used in order to interpret Eq.~\eqref{expoA} as a fluctuation relation. As discussed by Talkner {\it et al.} in Ref.~\cite{talkner}, only the joint probability distribution of the internal energy change of the system and heat exchanged with the environment satisfy proper fluctuation relations, i.e., equalities between quantities that do not depend on the total process but only on its end points. While in order to evaluate the left-hand side of Eq.~\eqref{expoA} we used the probability distribution for the heat exchanged with the environment in Eq.~\eqref{Qdist}. Nevertheless, it has been shown that relations which bear mathematical resemblance to fluctuation relations, can be derived for unital or bistochastic processes~\cite{kafri, rastegin, albash, rastegin2, johnkavan2}. A process is unital if and only if $\A = \openone_\e$, and in general a quantum operation is not unital. In the case considered here the process at hand is surely not unital, as the erasure of $\s$ would inevitably perturb a hypothetically prepared maximally mixed state of the environment (i.e., a Gibbs state at infinite temperature), therefore violating the condition that defines unitality of a process. The fluctuation-like relation in Eq.~\eqref{expoA} links the non-unital nature of the process being considered to the heat exchanged with the environment through the average of the function $e^{-\beta Q}$. We would also like to point out that, although a suggestive similarity exists between the form of Eq.~\eqref{expoA} and fluctuation relations derived under the explicit consideration of feedback mechanisms~\cite{sagawa,tasaki}, our study is very far from any feedback-based formalism. 

Operator $\A$ depends on the choice of states $\hat\rho_\s$. For practical applications one would be interested in computing the heat generated in erasing a state of choice. This would require computing $\A$ for each instance. On the other hand, given a thermal state $\hat\rho_\e$ and a unitary interaction $\hat U$, we can determine $\M$ as defined below Eq.~\eqref{expoM}. This is easily achievable by running a given circuit implementing $\hat U$ in reverse order, i.e., by taking the Hermitian conjugate of each elementary gates into which $\hat U$ is decomposed. Once $\M$ is determined, a non-trivial lower bound on heat generation is computed for any state $\rho_\s$ using Eq.~\eqref{boundtotal}. In this Letter we are working with the protocol used in Ref.~\cite{reeb} to derive a stronger version of the Landauer bound. The structure of our bound in Eq.~\eqref{boundtotal} stems solely from the exponentiated average heat, which relies on the information encoded in the full distribution rather than just the average. In the next section we will show, using specific physical models, that the bound derived in this Letter can be tighter than previously discovered bounds.

 \begin{figure}[b]
 \includegraphics[width=0.7\columnwidth]{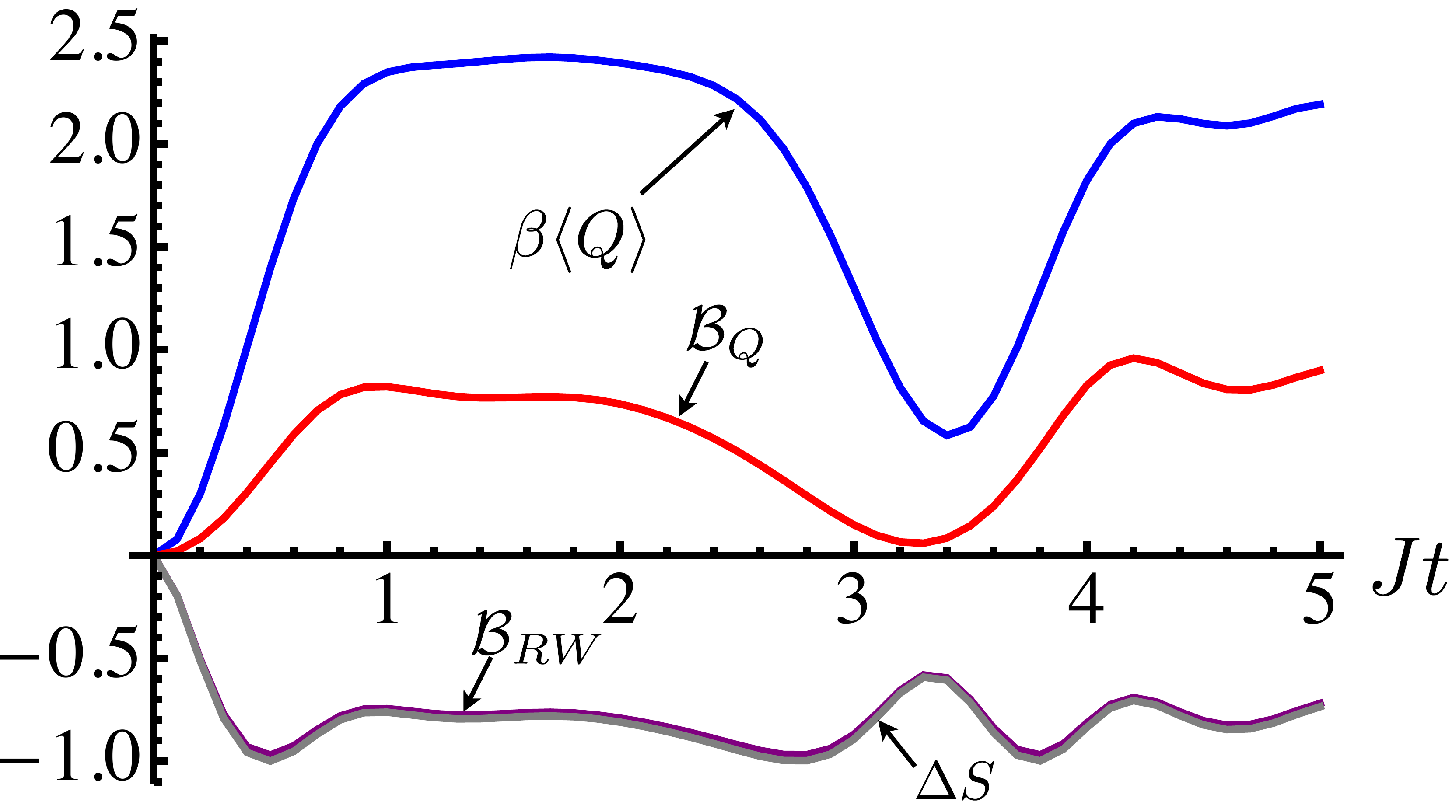}
 \caption{Similarly to Fig.~\ref{spinenvironment} {\bf (a)} and {\bf (b)}, we plot the key quantities of our study for $J_0/J=B_0/J=B/J=1$, $\beta=1$, $\alpha=1$ and an environment of $N=4$ elements. The curve showing the behaviour of $\Delta \ent$ is basically indistinguishable from the one for ${\cal B}_{RW}$.}
 \label{large}
 \end{figure}

{\it Physical model.---}
We consider a spin-1/2 system, whose logical states are labelled as $\ket{0}_\s$ and $\ket{1}_\s$, interacting with a environment embodied by an interacting spin chain of $N$ elements that we assume in contact with a thermostat at inverse temperature $\beta$. The system is sketched for illustrative purposes in Fig.~\ref{spinenvironment} {\bf (a)}. The Hamiltonian of the environment is given by the isotropic XX model
\begin{gather}
\hat H_\e = J \sum^{N-1}_{j=1} (\hat\sigma^j_x \hat\sigma^{j+1}_x + \hat\sigma^j_y \hat \sigma^{j+1}_y ) + B \sum^N_{j=1}\hat\sigma^j_z
\end{gather}
with $J$ the inter-spin coupling strength, $B$ the strength of the coupling between each spin and a homogeneous external magnetic field and $\hat\sigma^j_k$ is the $k$-Pauli spin operator ($k=x,y,z$) for particle $j=1,..,N$. The system spin is coupled to the environment through element $1$ of the chain according to the model
\begin{gather}
\hat H_{\se}=J_0\sum_{k=x,y}\hat\sigma^\s_k\hat\sigma^1_k.
\end{gather}
The Hamiltonian of the system spin in the absence of an environment is $\hat H_{\s}=B_0\hat\sigma^\s_z=B_0(\ket{1}\bra{1}_\s-\ket{0}\bra{0}_\s)$. In line with the general requirements of the erasure protocol highlighted before, we will assume that the environmental system is prepared in the equilibrium state $\rho_\e=e^{-\beta\hat H_\e}/{\cal Z}_\e$. For $B_0=B$ and $J_0=J$, the dynamics resulting from the model $\hat H=\hat H_\s+\hat H_\e+\hat H_{\se}$ is then equivalent to that of a time-dependent generalised amplitude damping channel described by the Kraus operator 
\begin{align}
& \hat A_{0}=\sqrt{p}\begin{pmatrix}\phi&0\\0&1\end{pmatrix},~~\hat A_{1}=\sqrt{p}\begin{pmatrix}0&0\\\sqrt{1-\phi^2}&0\end{pmatrix}, \\
& \hat A_{2}=\sqrt{1-p}\begin{pmatrix}1&0\\0&\phi\end{pmatrix},~~\hat A_{3}=\sqrt{1-p}\begin{pmatrix}0&\sqrt{1-\phi^2}\\0&0\end{pmatrix}
\notag
\end{align}
with $\phi={}_\s\bra{1}_\e\bra{0\cdots0}e^{-i\hat H t} \ket{0\cdots0}_\e \ket{1}_\s$~\cite{bose} and $p\in[0,1]$ a probability whose value is linked to the equilibrium temperature of the environment. Needless to say, the explicit form of $\phi$ depends on the size of the environmental subsystem, i.e. on the number $N$ of spins that compose it. Notice that the generalised amplitude damping channel addressed here is non-unital, and is thus perfectly suited to illustrate our main result.
 
We allow us the freedom to prepare the system spin in any pure state $\ket{\psi_i}_\s=\alpha\ket{1}_\s+\sqrt{1-\alpha^2}\ket{0}_\s$ with $\alpha\in{\mathbb R}$, for simplicity (the generality of our results is not affected by such assumptions, and the case of an arbitrarily mixed system state can be equally considered). In this case, the process at hand will make the entropy of the system increase. We start addressing the case of a single-spin environment, which provides a useful benchmark for our analysis, and then attack the case of a larger-size $\e$ to investigate the scaling of our findings with the dimension of the environmental Hilbert space. In such simple, yet interesting case, we easily get $\phi=\cos(2J t)$ and $p=e^{-\beta B}/{\cal Z}_\e=[1+\tanh(\beta B)]/2$, which in turn lead us to 
\begin{align}
& \beta \braket{\q} = B\sin^2(2J t)(2\alpha^2+\tanh(\beta B)-1),\\
& {\cal B}_\q = -\ln\left[2 (1-\phi ^2)(\alpha^2+p-2p\alpha^2)+\phi ^2\right]
\end{align}
with $\hat\rho'_\s=\tr_\e [e^{-i\hat H t} \ket{\psi_i} \bra{\psi_i}_\s \otimes \hat\rho'_\e e^{i\hat H t}]$ being the reduced state of the system $\s$ at time $t$. We now use the definition $\Delta \ent = \ent (\hat\rho_\s)- \ent (\hat\rho'_\s)$ with $\ent (\hat\rho) = -\tr [\hat\rho \ln(\hat\rho)]$ the von Neumann entropy of a generic state $\hat\rho$. As $\ent[\ket{\psi_i}\bra{\psi_i}_\s]=0$, we have $\Delta \ent= - \ent[\hat\rho'_\s]$. As this quantity turns out to be negative at all times, the relevant bound from Ref.~\cite{reeb} takes the form
\begin{gather}
{\cal B}_{RW}={\cal R}-\sqrt{{\cal R}^2+2 {\cal R} \ent [\hat\rho'_\s]}
\end{gather}
with ${\cal R}={\displaystyle\max_{0<r<0.5}}r(1-r)\ln^2[(1-r)(d-1)/r]$ and $d=2^{N-1}$.
Although $\ent [\hat\rho'_\s]$ can be computed analytically, its expression is too cumbersome to be reported here. A comparison among such quantities is made in Fig.~\ref{spinenvironment} {\bf (b)} and {\bf (c)}. The first of such plots shows that, at a set value of the environmental temperature, ${\cal B}_\q\ge{\cal B}_{RW}$ when $\alpha\simeq1$, i.e. when the system preparation is close to the fully polarised state $\ket{1}_\s$. This persists against temperature: in Fig.~\ref{spinenvironment} {\bf (c)} we present the behavior of $\beta \braket{\q}$, ${\cal B}_\q$, and ${\cal B}_{RW}$ for $\beta=10$. Needless to say, although we have demonstrated that the non-unital nature of the process at hand can indeed be advantageous, this feature is {\it per se} not sufficient to guarantee that  our bound to the amount of heat exchanged with the environment is tighter than ${\cal B}_{RW}$ for any value of the parameters at hand. Indeed, there are regions in the $(Jt,\alpha,\beta)$ space where the dimension-dependent quantity proposed in Ref.~\cite{reeb} gives a tighter bound to $\beta\braket{\q}$ than ${\cal B}_\q$.

Finally, it is worth addressing the case of larger environmental systems to study the effects of dimensionality. To this end, we have considered the fully isotropic model for $N$ growing from $1$ to $6$, finding results that are in agreement with the analysis presented above, albeit the interacting nature of the environmental subsystem makes the dynamics much richer than in the simple case of $N=1$, as shown in Fig.~\ref{large} for the case of $N=4$. Evidently, as the size of the environment grows $\Delta \ent$ and ${\cal B}_{RW}$ become basically indistinguishable. While this study is sufficient to illustrate the key features of the model at hand, a more extensive analysis of the implications of criticality and the full assessment of size-scaling will be presented elsewhere. 

{\it Conclusions.---}
Within the framework of the quantum distribution of heat we have proposed a framework for the addressing of Landauer's erasure principle. Our approach provides a bound to the amount of heat exchanged with a finite-size environment during the erasing process. Using the average of the exponential of the heat exchanged, our study links explicitly the non-unitality of a map to the erasure process and provides an example of how such feature might be exploited to provide a better bound to the amount of exchanged heat. We have illustrated such possibility by addressing the case of a generalised amplitude damping channel as the erasure mechanism for a single spin-1/2 system. Our study opens up a number of interesting avenues linked, for instance, to the study of the heat exchanges in microscopic models for system-environment interaction, the link to the emergence of non-Markovian features, design of quantum enhanced transport networks and the performance of thermodynamical cycles in quantum-inspired machines and motors~\cite{kosloff, sta}.

{\bf Acknowledgments.} We are grateful to T. J. G. Apollaro, L. C\'eleri, G. De Chiara, L. Mazzola, D. Reeb, J. Anders, R. Sarthour, and V. Vedral for invaluable discussions and suggestions. We acknowledge financial support from the Alexander von Humboldt Foundation, the UK EPSRC, 
the John Templeton Foundation (grant IDs 21806 and 43467), and the EU Collaborative Project TherMiQ (Grant Agreement 618074). This work was partially supported by the COST Action MP1209.


\end{document}